\documentclass[prd,twoside,showpacs,aps,longbibliography,twocolumn]{revtex4-2}
\usepackage{amsmath}
\usepackage{amssymb}
\usepackage{amsfonts}
\newcommand{\vecg}{\boldsymbol}
\renewcommand{\vec}{\textbf}
\newcommand{\ket}[1]{|#1\rangle}
\newcommand{\bra}[1]{\langle#1|}
\newcommand{\bracket}[2]{\langle#1|#2\rangle}
\newcommand{\sg}[1]{\mathsf{#1}}
\newcommand{\amp}[1]{\mathsf{#1}}

\DeclareMathOperator{\tr}{Tr} 
\usepackage{subfig}
\usepackage{graphicx}
\usepackage{color}
\begin{document}

\title{Quantum Field Theory of Space-like Neutrino}

\author{Jakub Rembieli\'nski}
\email{jaremb@uni.lodz.pl}
\author{Pawe{\l}{} Caban}
\email{P.Caban@merlin.phys.uni.lodz.pl}

\affiliation{Department of Theoretical Physics,\\
	Faculty of Physics and Applied Informatics, University of {\L}{\'o}d{\'z}\\
	Pomorska 149/153, PL-90-236 {\L}{\'o}d{\'z}, Poland}

\author{Jacek Ciborowski}
\email{cib@fuw.edu.pl}
\affiliation{Department of Physics, University of Warsaw\\
	Pasteura 5, PL-02-093 Warsaw, Poland}

\begin{abstract}
We performed a Lorentz covariant quantization of the spin-1/2 fermion field assuming 
the space-like energy-momentum dispersion relation. 
We a\-chieved the task in the following steps: 
($i$) determining the unitary realizations of the inhomogenous Lorentz group 
in the preferred frame scenario by means of the Wigner-Mackey induction procedure 
and constructing  the Fock space; 
($ii$) formulating  the theory in a manifestly covariant way by constructing 
the field amplitudes according to the Weinberg method; 
($iii$) obtaining the final constraints on the amplitudes by postulating 
a Dirac-like free field equation. 
Our theory allows to predict all chiral properties of the neutrinos, 
preserving the Standard Model dynamics. 
We discussed the form of the fundamental observables, energy 
and helicity, and show that  non-observation of the $+\tfrac{1}{2}$ helicity state 
of the neutrino and the  $-\tfrac{1}{2}$ helicity state of the antineutrino could 
be a direct consequence of the ``tachyoneity'' of neutrinos at the free level.
We found that the free field theory of the space-like neutrino is not invariant 
under the C and P transformations separately but is CP-invariant. 
We calculated and analyzed the electron energy spectrum in tritium decay 
within the framework of our theory and found an excellent agreement with the 
recent measurement of KATRIN.
In our formalism the questions of negative/imaginary energies and the causality 
problem does not appear. 
%\keywords{Space-like neutrino \and Second keyword \and More}
%\PACS{PACS code1 \and PACS code2 \and more}
\end{abstract}

\maketitle

\section{Introduction}

After nearly 90 years of the neutrino history this particle is still an enigma 
with a number of unanswered questions in the neutrino physics and the Standard Model.
We know that neutrinos oscillate so at last two neutrino generations 
are massive. 
Therefore the neutrino field should  possess two spin components and consequently
neutrinos should be found in two helicity states. 
However, only the left-handed helicity component 
of the neutrino and the right-handed of the antineutrino have been observed in experiments. 
A standard but rather technical explanation of this fact makes use of the 
see-saw mechanism \cite{MS1980_Neutrino-mass_PRL44.912,MW1980_Neutrino-mass,%
SV1980_Neutrino-mass_PRD22.2227}. 
In contrast,
we adopt a hypothesis that the neutrino is a particle satisfying 
the space-like dispersion relation. 
This assumption is suggested by a repeating occurrence of negative values for the neutrino
mass squared measured in numerous recent tritium-decay experiments 
\cite{Aker_KATRIN2019_PhysRevLett.123.221802,%
aker2019_KATRIN_tritium_first,Weinheimer_etal1999,Lobashev_etal1999_Troitsk,%
Kraus_etal2005_Mainz,Aseev_etal2011_Troitsk_PhysRevD.84.112003}.
This observation does not make a proof that neutrino is a tachyon because of an insufficient 
level of confidence of each of these separate results, 
however, it encourages considering
a possible theoretical descriptions of this possibility. 
Such trials have already been undertaken in the past. 
A hypothesis that neutrino might be a space-like particle was first discussed 
by Chodos et al. \cite{ChHK1985_neutrino-tachyon}.
Some arguments supporting this proposition were also presented 
by Giannetto et al. \cite{GMMR_1986}.
However, these attempts were unsuccessful due to the fact
that the standard relativistic quantum 
field theory is inapplicable for describing space-like particles, as pointed out by 
Kamoi and Kamefuchi \cite{KK1971_QTF-tachyons} and Nakanishi \cite{Nakanishi_1972_QFT_indef-metric}.
On the other hand, it was shown \cite{cab_Rembielinski1997} that an approach 
involving the notion of a preferred frame (PF) allows to construct a Lorentz-covariant 
quantum field-theoretical model of a relativistic helicity-$\tfrac{1}{2}$ fermionic tachyon.
In this way
one can avoid the fundamental difficulties related to the lack of a finite 
lower energy limit, appearance of infinite spin multiplets and causal problems 
appearing in the standard attempts to describe tachyons. 
In the space-like neutrino case
the Cosmic Neutrino Background (CNB) frame, an artefact of the electroweak phase 
transition \cite{Baumann_etal2019_neutrino},
is a natural candidate for the preferred frame. 
In the above framework the $\beta$ decay
was considered in \cite{CR1999_beta_decay_tachyonic_neutrino}
and the corresponding decay rate (energy spectrum)
for electrons was derived and discussed 
in the context of the neutrino mass measurement. 
For other contributions to the tachyonic neutrino hypothesis see 
\cite{CRS2006_Decays_spacelike_neutrinos,Radzikowski2010-tachyonic-neutrino,%
CRSW_2006_Oscillations,ChKPG1992_neutrino,Ehrlich2015_Neutrino-tachyon}.

References to the notion of a PF in the context of the quantum theory are owed 
to several authors. 
The historical term of ``aether'' used in the field--theoretical context by Dirac 
\cite{Dirac1951_aether,Dirac1951_electron}
was superseded by that of the PF, as, e.g., in de Broglie–Bohm formulation 
of quantum mechanics \cite{cab_Ghost_in_atom1986,Bohm1952_I}.
Bell suggested that it would have been helpful to consider a PF at the fundamental level 
for resolving incompatibilities between special relativity theory and nonlocality of 
quantum mechanics \cite{cab_Bell1981}---an opinion also shared by other authors 
\cite{Gisin2014-in-book_correl-Newtonian,cab_ZBGT2000,cab_CR1999,%
RC2018-PhysRevA.97.062106,cab_RS2002,KRS2007-PhysRevD.76.045018}.
Let us mention in particular: Lorentz-violating extensions of the Standard Model
\cite{CK1997-PhysRevD.55.6760,CK1998-PhysRevD.58.116002,CG1999-PhysRevD.59.116008,%
CG2006-PhysRevLett.97.021601}, approaches to classical and quantum gravity 
like the Einstein's aether \cite{Jacobson2008} and Ho\v{r}ava-Lifshitz theories of gravity
\cite{Horava2009-gravity} (including vacuum solutions in this model
\cite{Rembielinski2014-Horava}), and the so-called doubly special relativity theories
\cite{Amelino-Camelia2013}, characterized by modified dispersion relations, 
common for the Lorentz violating models. 
In almost all of the above theories specific physical effects are predicted, 
of magnitude usually suppressed by a power of the Planck scale, like, e.g., vacuum birefringence
\cite{CK1997-PhysRevD.55.6760,CK1998-PhysRevD.58.116002,CG1999-PhysRevD.59.116008,%
CG2006-PhysRevLett.97.021601,Jacobson2008,Amelino-Camelia2013}. 
This brief outlook demonstrates that the concept of a PF has been frequently 
referred to in the context of Lorentz symmetry violation within numerous contemporary 
theories.

It is important to stress that if tachyons exist, one could,  in principle, 
consider synchronizing distant clocks in almost absolute way 
(instantaneous synchronization) in the limit of zero energy (infinite velocity) 
if they interact with matter with finite probability in these conditions.
In consequence, one has a possibility of introducing absolute time between observers.
The only way to reconcile this implication with the Lorentz symmetry 
lies in an assumption of existence of a preferred frame \cite{cab_Rembielinski1997}. 
We discuss this point in Sec.~\ref{sec:tach-kinematics}.

In the present paper we formulate a fully consistent quantum field theory of the space-like
neutrino with both one-half helicity components under the assumption of the existence of
the preferred frame. 
In our approach the states of tachyonic field excitations observed from an arbitrary 
inertial frame  should depend on the PF four-velocity as seen from the observer frame. 
All other physical fields/states are simply unaffected by the PF.
This means that the PF is ralated with the tachyonic sector only.
As we will see
the preferred frame concept is crucial for the successful construction of a viable 
theory of fermionic tachyons enabling to eliminate the issue of negative/i\-ma\-ginary 
energies and causality problems as well as to perform the field quantization procedure 
in arbitrary inertial frames. At the interaction level we obtain the anomaly free, 
perturbatively renormalizable variant of the electroweak model.
We construct this theory in few steps. 
Firstly, 
in Sec.~\ref{sec:tach-kinematics} we construct the one-particle phase space 
and discuss its Lorentz covariance 
as well as the issue of imaginary energies, stability of the theory and causality.
Next, in Sec.~\ref{sec:space-of-states},
we determine the unitary realizations of the inhomogenous Lorentz group 
under the condition of existence of an inertial preferred frame. 
We do this by means of the Wigner-Mackey induction procedure \cite{Mackey1968}.
Secondly, in Sec.~\ref{sec:Fock-space}, we construct the Fock space 
by generation of the multiparticle basis states
from the Lorentz invariant vacuum state by using the corresponding algebra of 
creation/annihilation operators. 
This quantization procedure is formulated for all inertial frames.
Thirdly, in Sec.~\ref{sec:manifestly-cov-form}, we formulate the theory in a manifestly 
covariant way by constructing the field amplitudes according to the Weinberg 
method \cite{cab_Weinberg1964}.
Fourthly, in Sec.~\ref{sec:Dirac-like-eq}, we obtain the final constraints 
on the amplitudes by postulating a Dirac-like free 
field equation determining the free space-like neutrino field completely. 
Next, in Sec.~\ref{sec:helicity}
we show that the amplitudes of the
space-like neutrino field have properties which 
are consistent with the high energy chiral properties  of  the neutrino observed in reality.
On the other hand, we predict that at very low neutrino energies, 
both helicities should be observed.
We discuss the form of the fundamental observables, namely energy and helicity. 
In Sec.~\ref{sec:discrete-sym}
we find that the free field theory of the space-like neutrino is not invariant under 
the parity and charge conjugation transformations separately, however, 
it is invariant under the CP transformation. 
Finally, in Sec.~\ref{sec:beta-decay} we construct the anomaly free variant 
of the electroweak  model and
calculate and analyze the $\beta$ decay 
rate (electron energy spectrum) in this context. 

Because the tachyon kinematics in the preferred frame scenario as well as 
the construction of the corresponding quantum field theory is not widely known, 
we present our approach with necessary details, restricting ourselves to one neutrino 
generation for simplicity.

\section{Tachyon kinematics in the preferred frame scenario}
\label{sec:tach-kinematics}

Tachyon is a particle with a space-like four-momentum 
$k^\mu$, satisfying the following Lorentz-covariant dispersion relation
\begin{equation}
k^\mu k_\mu = {k^{0}}^2 - \vec{k}^2 = - \kappa^2,
\label{dispersion_tachyon}
\end{equation}
where $\kappa$ denotes the ``tachyonic mass'' ($\kappa>0$).
However, $\kappa$ can be also viewed as a residual momentum of the tachyon 
in the limit of zero energy.
Eq.~(\ref{dispersion_tachyon}) defines a one sheet hyperboloid 
where energy $k^0$ takes the values from minus to plus infinity.
Lorentz transformations relating inertial frames can transform positive 
$k^0$ to any negative value. Hence, this causes 
the vacuum instability on the quantum level, i.e., 
possibility of a spontaneous creation from the vacuum of
pairs of particles with the total four-momentum equal to zero. 
Below we show that this problem can be resolved
in the preferred frame scenario.
Let us denote the inertial observer's frame by $\Sigma_u$. 
From his point of view the PF, $\Sigma_{PF}$, moves with a constant four-velocity
$u^\mu$ satisfying
\begin{equation}
{u^0}^2 - \vec{u}^2 = 1.
\label{u_square}
\end{equation}
Thus for observers stationary in the PF,
its four-velocity is given by $u_{PF}=(1,0,0,0)$.
The PF three-velocity (in units of $c$) is given by  
$\vec{V}=\vec{u}/u^0$ so $u^0={1}/{\sqrt{1-\vec{V}^2}}$
is the Lorentz factor of the PF. 
Now, using the four-momentum $k^\mu$ of the tachyon and the four-velocity  
$u^\mu$ we can construct a Lorentz invariant
\begin{equation}
q = uk = u^\mu k_\mu = u^0 k^0 - \vec{u}\cdot\vec{k}
\label{q_def}
\end{equation}
equal to the tachyonic energy measured in the preferred frame.
Hence, the physically acceptable four-momenta $k^\mu$ are bounded 
by the Lorentz covariant condition
\begin{equation}
q>0,
\label{q_greater_zero}
\end{equation}
guaranting nonnegativity of energy in the preferred frame.
We see from Eq.~(\ref{q_def}) that only in the preferred frame, $\Sigma_{PF}$,
tachyons have always non negative energies. 
Indeed,  applying the Lorentz boost transformation one can also obtain kinematical 
states of the tachyon with negative energies in other inertial 
frames 
(but still with a lower bound set by Eq.~(\ref{q_greater_zero})).
The explicit form of the  energy and the value of the momentum obtained 
from the relations 
(\ref{dispersion_tachyon},\ref{u_square},\ref{q_def},\ref{q_greater_zero}) 
in an arbitrary inertial frame $\Sigma_u$ is the following 
\begin{multline}
k^0=\\
\frac{q u^0 + \cos\theta \sqrt{(u^0)^2-1}%
\sqrt{q^2+\kappa^2[\cos^2\theta+(u^0)^2\sin^2\theta]}}%
{\cos^2\theta + (u^0)^2 \sin^2\theta},
\label{k0_q}
\end{multline}
\begin{multline}
|\vec{k}| \equiv \omega(q,u,\theta)  = \\
\frac{q \cos\theta \sqrt{(u^0)^2-1}%
+u^0\sqrt{q^2+\kappa^2[\cos^2\theta+(u^0)^2\sin^2\theta]}}%
{\cos^2\theta + (u^0)^2 \sin^2\theta},
\label{omega_k_u}
\end{multline}
where $q>0$,  $u^0={1}/{\sqrt{1-\vec{V}^2}} $,
$\theta$ is the angle between the momentum,
$\vec{k}$, and the PF velocity, $\vec{V}$. 
Thus, the energy bound in an arbitrary frame $\Sigma_u$ has the following form 
\begin{equation}
k^0 > \frac{\kappa |\vec{V}|
	\cos{\theta}}{\sqrt{1-(|\vec{V}|\cos{\theta})^2}}.
\label{k0_theta}
\end{equation}

We introduce an invariant measure respecting the dispersion
relation (\ref{dispersion_tachyon}) and the covariant condition
(\ref{q_greater_zero}): 
\begin{align}
d\mu(k,u) & = d^4 k \delta(k^2+\kappa^2) \Theta(uk) \nonumber\\
& = \frac{1}{2u^0} |\vec{k}| \, 
\Theta(q) \, dq \,d\Omega,
\label{measure_inv}
\end{align}
(see \ref{app:measure}),
where $d\Omega$
is the solid angle differential related to the neutrino momentum $\vec{k}$,
$|\vec{k}|$ is given in Eq.~(\ref{omega_k_u}) and $\Theta$
is the Heaviside Theta function. 

Now, we comment common objections against the tachyonic theories in the context 
of the formalism introduced above.\\
\textit{Problem of negative energies}\\
Negative energies can formally appear in the standard description of tachyons 
for the reason that covariant lower bound of energy does not exist in this case. 
In contrast, existence of a preferred frame allows to define the Lorentz covariant 
condition (\ref{q_greater_zero}), constituting the lower bound of energy in each 
inertial frame, eliminating possibility of the kinematical instability. 
Indeed, if  the four-momentum  $k^\mu$ belongs to the upper half of the 
one-sheet hyperboloid, i.e., it satisfies the inequality $q=uk>0$ then
$-k^\mu$ must belong to the lower part of hyperboloid because
$(-uk) = -q<0$ does not satisfy the condition (\ref{q_greater_zero}). 
Therefore, the condition (\ref{q_greater_zero}) eliminates the possibility 
of the kinematical instability. 
We note that this condition is analogous to choosing the upper energy-momentum 
cone for massless particles as the physical one.\\
\textit{Problem of imaginary  energies}\\
This is rather fictitious problem in the case of particles satisfying the dispersion 
relation (\ref{dispersion_tachyon}). It is rooted to the fact that 
Eq.~(\ref{dispersion_tachyon}) implies 
$k^0 \sim \sqrt{\vec{k}^2 - \kappa^2}$ so $k^0$
can be in principle imaginary for $\vec{k}^2 < \kappa^2$ and we can expect 
exponentially divergent trajectories in that case. However this is impossible 
for tachyons on the mass shell, i.e., satisfying the dispersion relation 
(\ref{dispersion_tachyon}) because in that case the inequality
$|\vec{k}|>\kappa$ must hold for physical momenta. 
In our formalism this is guaranteed by the choice of the invariant measure 
(\ref{measure_inv}) which vanishes inside the sphere defined by inequality 
$\vec{k}^2 = \kappa^2$. 
Absence of the imaginary eigenvalues of the energy operator is also evident 
in the explicit formulation of our theory in 
Secs.~\ref{sec:Fock-space}--\ref{sec:Dirac-like-eq}.\\
\textit{Problem of non-causal behavior}\\
The notion of causality is inseparably related to the definition of the 
coordinate time. In the standard relativistic theories time is identified with 
the zeroth coordinate in the Minkowski space-time. 
Consequently, for a space-like separation of events, the Lorentz transformations 
can change their time ordering. However, existence of a preferred frame provides 
a solution of this problem. 
Namely, we can introduce another, Lorentz invariant, dynamical parameter, 
$T:= u_\mu x^\mu$. Notice, that this parameter allows a Lorentz invariant 
$T$-time ordering is in this case Lorentz invariant.
In the PF, $T=x^{0}_{PF}$, i.e., $t_{PF}=T$
so it is equal to the Einsteinian time (in the $c$ units).
In an arbitrary frame $\Sigma_u$
from the definition of $T$, we obtain
$\tau\equiv\tfrac{t_{PF}}{u^0}=t-\vec{V}\cdot\vec{x}$
where $\vec{V}$ is the defined above velocity of the preferred frame, i.e., 
$\vec{V}=\vec{u}/u^0$. The coordinate time redefinition 
$\tau=t -\vec{V}\cdot\vec{x}$
between the Einsteinian time $t$ and the time $\tau$ is simply the admissible 
change of clock synchronization \cite{cab_Rembielinski1997}.
For a stationary observer, in each fixed point ($d\vec{x}=0$) $dt=d\tau$, 
i.e., the flow of the time is the same. 
However, it follows from the time redefinition formula that
$\tfrac{d\tau}{dt}=1-\vec{V}\cdot\vecg{\vartheta}$,
where the velocity $\vecg{\vartheta}=\tfrac{d\vec{x}}{dt}$.
Notice that for subluminal or luminal motion  ($|\vecg{\vartheta}|\le1$) 
the derivative  $d\tau/dt$ is always nonnegative, i.e., the arrows of the Einsteinian 
time $t$ and the time $\tau$ are the same (causality, i.e., sequence of events is the same) .  However, for superluminal motion
($|\vecg{\vartheta}|>1$), the derivative $d\tau/dt$ can change sign to negative 
because the scalar product $\vec{V}\cdot\vecg{\vartheta}$ can, in some 
velocity configurations, exceed 1. 
Because $d\tau$ is always positive (the arrow of time $t_{PF}$ is fixed) then $dt$ 
must change the sign, which corresponds to the indeterminacy of  the causal relation 
in the Einsteinian synchronization scheme for velocities higher than the light velocity while, 
in terms of the time $\tau$, it is determined for all velocities. 
Moreover, the time-$\tau$ synchronization satisfies the crucial physical 
requirement---the average value of the light velocity over closed paths 
is frame independent and equal to $c$. 
For more information on the issue of clock synchronization in the special relativity see 
\cite{cab_AVS1998,Lammerzahl2005_SR-Lorentz-inv,Jammer2006}.
Summarizing, only the $\tau$ synchronization is adequate to determine 
uniquely the sequence of events (causality) for phenomena with participation of tachyons 
in the presence of the PF or equivalently, causality should be referred
to observers stationary in the PF.

\section{Space of states of the space-like neutrino}
\label{sec:space-of-states}

The condition \textit{sine qua non} to apply the tachyonic hypothesis to neutrino physics 
is to construct a quantum field theory of space-like fields. 
Indeed, the standard relativistic field theory is inapplicable to this case because of two reasons:
Firstly, the Wigner little group of the space-like four-momentum of the tachyon is the 
noncompact $\sg{SO}(2,1)$ group, so its unitary representations are 
scalar or infinite-dimensional. 
Consequently, the spin multiplets are either one-dimensional or infinite-dimensional.
Therefore, a spin-$\tfrac{1}{2}$ neutrino cannot be identified with such representations. 
Secondly, if we go around the above problem and try describing the tachyonic neutrino with 
the help of the bi-spinor nonunitary representation by identifying the Lagrange density
appropriate to the space-like dispersion relation as is done, e.g., in
\cite{ChHK1985_neutrino-tachyon}, we evidently loose unitarity, 
i.e., the probabilistic character of the quantum description.  
However, in the scenario with a PF, these difficulties do not arise.

As was stated above, a reasonable description of tachyons needs the presence of a
preferred frame. Consequently, the tachyonic neutrino basis states  should be dependent 
not only on the space-like four-momentum, $k^\mu$, but also on the PF four-velocity,  
$u^\mu$. Our aim is to apply the Wigner-Mackey induction procedure in this case. 
Hereafter we will denote eigenvectors of the four-momentum operator as 
$\ket{k,u,\lambda}$, where $\lambda$ is identified with the
neutrino/antineutrino helicity. 
Now, an observer in an arbitrary reference frame $\Sigma_u$ 
has at his disposal the Hilbert space  
$\mathcal{H}_u$ of neutrino states. 
The family of Hilbert spaces $\mathcal{H}_u$ form a bundle corresponding to 
the bundle of inertial frames $\Sigma_u$, i.e., to the quotient manifold 
$\sg{SO}(1,3)/\sg{SO}(3)\sim {\mathbb{R}}^3$ as the base space and 
with $\mathcal{H}_u$ as fibers.
To apply the Wigner-Mackey construction to this case we should find the little 
group of the pair of four-vectors $(k,u)$ determining the neutrino state. 
To do this we transform the pair $(k,u)$  to the preferred frame by action of the boost 
$L_u^{-1}$ to obtain the pair $(L_u^{-1} k=k^\prime,L_u^{-1} u=u_{PF})$.
Next we rotate $k^\prime$ with the help of the rotation $R_{\vec{n}}^{-1}$ 
to the $z$-axis in order
to obtain $\tilde{k}=(L_u R_{\vec{n}})^{-1} k=(q; 0,0,\sqrt{q^2+\kappa^2})$ 
while $u_{PF}$ is left unchanged 
(for the explicit form of  $L_u$  and $R_{\vec{n}}$ see \ref{app:rotations}).
Thus, the pair $(k,u)$ can be obtained from the standard pair $(\tilde{k},u_{PF})$
by a sequence of Lorentz transformations $L_u R_{\vec{n}}$, i.e.,
\begin{equation}
(k,u) = L_u R_{\vec{n}}(\tilde{k},u_{PF})
\label{k_u_generacja}
\end{equation}
provided the unit vector $\vec{n}$ is equal to
\begin{equation}
\vec{n} = \vec{n}(k,u) = 
\frac{1}{\sqrt{q^2+\kappa^2}} 
\Big(
\vec{k} - \frac{q+k^0}{1+u^0} \vec{u}
\Big),
\label{vec_n_def}
\end{equation}
where, by means of Eq.~(\ref{q_def}) and (\ref{q_greater_zero}),
$k^0=(q+\vec{u}\cdot\vec{k})/u^0$
and $q>0$.
It is obvious that the orthogonal group $\sg{SO}(2)$ is
the stability group of the pair $(\tilde{k},u_{PF})$. 
Therefore, irreducible unitary orbits of the inhomogeneous Lorentz 
group should be induced from the unitary irreducible representations of 
$\sg{SO}(2)$, i.e., from the one-dimensional representation of the $\sg{U}(1)$ group.
Applying the Wigner--Mackey procedure to the basis vectors
$\ket{k,u,\lambda}$, $\lambda\in{\mathbb{R}}$,
in the  manifold of Hilbert spaces ${\mathcal{H}}_u$  we obtain a result that the unitary 
action of the Lorentz group is of the form
\begin{equation}
U(\Lambda) \ket{k,u,\lambda} = 
e^{i\lambda\varphi(\Lambda,k,u)}
\ket{\Lambda k,\Lambda u,\lambda},
\label{U_Lambda_ket}
\end{equation}
where $e^{i\varphi(\Lambda,k,u)}$
is the phase factor corresponding to the Wigner rotation
\begin{equation}
W(k,u,\Lambda) = (L_{\Lambda u} R_{\vec{n}(\Lambda k,\Lambda u)})^{-1}
\Lambda L_u R_{\vec{n}(k,u)}.
\label{Wigner_rotation}
\end{equation}
The standard arguments lead to integer or half-integer values of $\lambda$.
In the case of the tachyonic neutrino we restrict further considerations to the values
$\lambda = \pm \tfrac{1}{2}$.

Moreover, in each fixed space ${\mathcal{H}}_u$  we adopt the following 
Lorentz covariant
normalization for the momentum eigenstates for neutrinos and antineutrinos
\begin{equation}
\bracket{k,u,\lambda}{k^\prime,u,\lambda^\prime} = 
\tfrac{1}{|\vec{k}|}
2 u^0 \delta_{\lambda\lambda^\prime} \delta(q-q^\prime) 
	\delta(\Hat{\vec{k}} - \Hat{\vec{k}}{}^\prime),
\label{normalization}
\end{equation}
where $\Hat{\vec{k}}$ denotes the unit vector ${\vec{k}}/{|\vec{k}|}$
and $\delta(\Hat{\vec{k}} - \Hat{\vec{k}}{}^\prime)$
is the spherical Dirac delta (see \ref{app:measure}).
Notice that it is convenient to choose the $z$-axis in the direction of 
the preferred frame velocity $\vec{V}$,
the angle $\theta$ is the polar angle of $\Hat{\vec{k}}$ in such a case.

Therefore, the one-particle space of tachyonic states is a direct sum
of one particle tachyonic neutrino $\oplus$ antineutrino 
space (of course neutrino and antineutrino spaces are mutually orthogonal).
The completeness relation in this space of has the form
\begin{equation}
\sum_\lambda \int d\mu(k,u) 
\Big(\ket{k,u,\lambda}_\nu\bra{k,u,\lambda} 
+ \ket{k,u,\lambda}_{\bar{\nu}}\bra{k,u,\lambda} \Big)
= I,
\label{completeness-relation}
\end{equation}
where the Lorentz invariant measure $d\mu$ is defined in Eq.~(\ref{measure_inv}).
We prove this relation explicitly in \ref{app:measure}.

\section{Fock space of the space-like neutrino}
\label{sec:Fock-space}

We construct a free field theory of the tachyonic neutrino in the preferred frame scenario, 
in close analogy with the standard formalism,
restricted to a single neutrino generation for simplicity. 
Because an irreducible realization of the Lorentz group is fixed in our case by 
the choice of 
$\lambda$, the basis vectors $\ket{k,u,\lambda}$
are obtained by the action of the neutrino creation operators $a_\lambda^\dagger(k,u)$ 
and antineutrino creation operators $b_\lambda^\dagger(k,u)$
on the normalized to the unity vacuum vector $\ket{0}$ defined  by the standard conditions
\begin{equation}
a_\lambda(k,u) \ket{0} = b_\lambda(k,u) \ket{0} = 0.
\label{vacuum}
\end{equation}
Thus
\begin{align}
\ket{k,u,\lambda}_\nu & = a_\lambda^\dagger(k,u) \ket{0},\\
\ket{k,u,\lambda}_{\bar{\nu}} & = b_\lambda^\dagger(k,u) \ket{0}.
\end{align}
The above vectors respect the normalization (\ref{normalization}) provided 
the creation and annihilation operators satisfy, for each fixed $u$, 
the following anti-commutation canonical relations
\begin{align}
&\{ a_\lambda(k,u),a_\sigma^\dagger(k^\prime,u)\}\nonumber\\
 & \phantom{a_\lambda(k,u)} = 
\frac{1}{\omega(q,u,\theta)} 2 u^0 \delta(q-q^\prime) 
	\delta(\Hat{\vec{k}} - \Hat{\vec{k}}{}^\prime) \delta_{\lambda\sigma},
\label{anticommutator_a}\\
&\{ b_\lambda(k,u),b_\sigma^\dagger(k^\prime,u)\}\nonumber\\
 & \phantom{a_\lambda(k,u)} =
\frac{1}{\omega(q,u,\theta)} 2 u^0 \delta(q-q^\prime) 
\delta(\Hat{\vec{k}} - \Hat{\vec{k}}{}^\prime) \delta_{\lambda\sigma}.
\label{anticommutator_b}
\end{align}
The remaining anti-commutators  vanish.

To reproduce formula (\ref{U_Lambda_ket}), 
creation operators should transform under the action of the Lorentz 
group according to the law
\begin{align}
U(\Lambda) a_\lambda^\dagger(k,u) U(\Lambda)^\dagger & 
= e^{i\lambda\varphi(\Lambda,k,u)} a_\lambda^\dagger (\Lambda k,\Lambda u),
\label{a_lambda_transf}\\
U(\Lambda) b_\lambda^\dagger(k,u) U(\Lambda)^\dagger & 
= e^{i\lambda\varphi(\Lambda,k,u)} b_\lambda^\dagger (\Lambda k,\Lambda u).
\label{b_lambda_transf}
\end{align}

The corresponding Fock space of multiparticle states can be now defined in a standard 
way by a succesive action of creation operators on the vacuum state..
At this stage, given the Lorentz invariant measure, the space of states and the 
corresponding transformation rules, we are ready to define the fundamental observables.
The helicity operator $\Hat{\lambda}$ is defined as
\begin{multline}
\Hat{\lambda}(u) = \sum_\lambda \int d\mu(k,u)
\lambda \Big(
a_\lambda^\dagger(k,u) a_\lambda (k,u)\\
+ b_\lambda^\dagger(k,u) b_\lambda (k,u)
\Big)
\label{lambda_def}
\end{multline}
and by means of (\ref{anticommutator_a}-\ref{anticommutator_b})
it satisfies
\begin{align}
[\Hat{\lambda},a_\lambda^\dagger(k,u)] & = \lambda a_\lambda^\dagger(k,u),
\label{commutator_lambda_a}\\
[\Hat{\lambda},b_\lambda^\dagger(k,u)] & = \lambda b_\lambda^\dagger(k,u).
\label{commutator_lambda_b}
\end{align}
Similarly, we define the covariant four-momentum operator 
$\Hat{P}_\mu$ by the standard formula
\begin{multline}
\Hat{P}_\mu(u) = \int d\mu(k,u) k_\mu \sum_\lambda
\Big(
a_\lambda^\dagger(k,u) a_\lambda (k,u)\\
+ b_\lambda^\dagger(k,u) b_\lambda (k,u)
\Big)
\label{P_def}
\end{multline}
which implies, that
\begin{align}
[\Hat{P}_\mu,a_\lambda^\dagger(k,u)] & = k_\mu a_\lambda^\dagger(k,u),
\label{commutator_P_a}\\
[\Hat{P}_\mu,b_\lambda^\dagger(k,u)] & = k_\mu b_\lambda^\dagger(k,u).
\label{commutator_P_b}
\end{align}

\section{Manifestly covariant formulation}
\label{sec:manifestly-cov-form}

The neutrino field $\nu(x,u)$ is defined as the Dirac bispinor operator of the form 
\begin{multline}
\nu_\alpha (x,u) = \frac{1}{(2\pi)^{3/2}}
\sum_\lambda  \int d\mu(k,u) 
\Big[
e^{ikx} \amp{v}_{\alpha\lambda}(k,u) b_\lambda^\dagger(k,u) \\
+ e^{-ikx} \amp{u}_{\alpha\lambda}(k,u) a_\lambda(k,u)
\Big],
\label{nu_x_u_Fourier}
\end{multline}
with the standard manifestly covariant transformation rule
\begin{equation}
U(\Lambda) \nu(x,u) U(\Lambda)^\dagger =
S(\Lambda^{-1}) \nu(\Lambda x,\Lambda u),
\label{nu_transformation}
\end{equation}
where $S(\Lambda)$ belongs to the representation
$D^{(\frac{1}{2},0)}\oplus D^{(0,\frac{1}{2})}$
of the homogenous Lorentz group.

In order to fulfill the transformation law
(\ref{nu_transformation}),
the amplitudes $\amp{v}_\lambda (k,u)$ and $\amp{u}_\lambda (k,u)$, 
must satisfy the Weinberg consistency conditions obtained with the use of
Eqs.~(\ref{a_lambda_transf},\ref{b_lambda_transf},\ref{nu_x_u_Fourier}),
namely
\begin{align}
\amp{u}_\lambda(\Lambda k,\Lambda u) &
= S(\Lambda) \amp{u}_\lambda(k,u) e^{-i\lambda\varphi(\Lambda,k,u)},
\label{Weinberg_u}\\
\amp{v}_\lambda(\Lambda k,\Lambda u) &
= S(\Lambda) \amp{v}_\lambda(k,u) e^{i\lambda\varphi(\Lambda,k,u)}.
\label{Weinberg_v}
\end{align}
In the following we choose the Weyl bi-spinor representation of the 
Lorentz group in the form
\begin{equation}
S(\Lambda(A)) = 
\begin{pmatrix}
A & 0 \\
0 & {A^\dagger}^{-1}
\end{pmatrix},
\label{S_Lambda}
\end{equation}
where the matrices $A$ belong to the $\sg{SL}(2,{\mathbb{C}})$ group.
The corresponding representation of the $\gamma$ matrices  is the following
\begin{gather}
\gamma^0=
\begin{pmatrix}
0 & I \\ I & 0
\end{pmatrix},\quad
\gamma^k=
\begin{pmatrix}
0 & \sigma^k \\ -\sigma^k & 0
\end{pmatrix},\\
\gamma^5 = \gamma_5 = 
i \gamma^0 \gamma^1 \gamma^2 \gamma^3 =
\begin{pmatrix}
-I & 0 \\ 0 & I
\end{pmatrix},
\end{gather}
where $\sigma^k$ are the standard Pauli matrices.
Note that the Pauli matrices are contravariant.
In the Weyl representation the neutrino field and
the amplitudes $\amp{v}(k,u)=[\amp{v}_{\alpha\lambda}(k,u)]$ and
$\amp{u}(k,u)=[\amp{u}_{\alpha\lambda}(k,u)]$ admit the following chiral
decompositions
\begin{equation}
\amp{u}(k,u)=\begin{pmatrix}
\amp{u}_L (k,u)\\ \amp{u}_R(k,u)
\end{pmatrix},\quad
\amp{v}(k,u)=\begin{pmatrix}
\amp{v}_L (k,u)\\ \amp{v}_R(k,u)
\end{pmatrix}
\end{equation}
corresponding to the chiral projections
$P_{L/R} = \tfrac{1}{2}(I\mp \gamma^5)$.

By means of Eqs.~(\ref{k_u_generacja},\ref{Wigner_rotation},\ref{Weinberg_u},%
\ref{Weinberg_v},\ref{S_Lambda}) we obtain
\begin{align}
\amp{u}(k,u) & = S(L_u R_{\vec{n}}) \amp{u}(\tilde{k},u_{PF}),
\label{amp_u_generacja}\\
\amp{v}(k,u) & = S(L_u R_{\vec{n}}) \amp{v}(\tilde{k},u_{PF}),
\label{amp_v_generacja}
\end{align}
where
\begin{equation}
S(L_u R_{\vec{n}}) = 
\begin{pmatrix}
\mathbb{L}_u \mathbb{U}_{\vec{n}} & 0\\
0 & \mathbb{L}_{u^\pi} \mathbb{U}_{\vec{n}}
\end{pmatrix},
\label{S_LR_def}
\end{equation}
and $\mathbb{L}_u$ is the Lorentz  boost matrix representing $L_u$
in the left-handed spinor space, $\mathbb{L}_{u^\pi}$ is the  boost acting 
in the right-handed spinor space, $\pi$ denotes the parity operation on 
four-vectors whereas $\mathbb{U}_{\vec{n}}$ represents the rotation 
$R_{\vec{n}}$. Explicitly
\begin{align}
\mathbb{L}_u & = \frac{1}{\sqrt{2(1+u^0)}}
\Big(  (1+u^0)I - \vec{u}\cdot\vecg{\sigma} \Big),
\label{Ln_explicit}\\
\mathbb{L}_{u^\pi} & = \mathbb{L}_{u}^{-1} = \frac{1}{\sqrt{2(1+u^0)}}
\Big(  (1+u^0)I + \vec{u}\cdot\vecg{\sigma} \Big),
\label{Ln_pi_explicit}\\
\mathbb{U}_{\vec{n}} & = \frac{1}{\sqrt{2(1+n^3)}}
\begin{pmatrix}
1+n^3 & -n^1+in^2\\
n^1+in^2 & 1+n^3
\end{pmatrix}
\label{Un_explicit}
\end{align}
and the unit vector $\vec{n}(k,u)$ is defined in Eq.~(\ref{vec_n_def}).

Now, choosing $k=\tilde{k}$, $u=u_{PF}$ in (\ref{Weinberg_u},\ref{Weinberg_v})
and $\Lambda$ as the stability group element of this pair, i.e.,
$\Lambda=R_z(\phi)$, where $R_z (\phi)$ is rotation around the third axis represented, 
according to (\ref{S_Lambda}), by the matrix
\begin{equation}
S(R_z(\phi)) = \begin{pmatrix}
\mathbb{U}_z(\phi) & 0 \\
0 & \mathbb{U}_z(\phi)
\end{pmatrix}
\end{equation}
with
\begin{equation}
\mathbb{U}_z(\phi) = 
\begin{pmatrix}
e^{i\phi/2} & 0 \\ 0 & e^{-i\phi/2}
\end{pmatrix}
\end{equation}
and noting that from (\ref{Wigner_rotation}) the Wigner phase is given by
$\varphi(R_z (\phi),\tilde{k},u_{PF}  )=\phi$ in this case, we obtain 
the following conditions for the chiral amplitudes 
\begin{align}
\mathbb{U}_z(\phi) \tilde{\amp{u}}_{\lambda_{L/R}}(\tilde{k},u_{PF}) 
e^{-i\lambda\phi} & = \tilde{\amp{u}}_{\lambda_{L/R}}(\tilde{k},u_{PF}),\\
\mathbb{U}_z(\phi) \tilde{\amp{v}}_{\lambda_{L/R}}(\tilde{k},u_{PF}) 
e^{i\lambda\phi} & = \tilde{\amp{v}}_{\lambda_{L/R}}(\tilde{k},u_{PF}),
\end{align}
leading to 
\begin{align}
\tilde{\amp{u}}_{-1/2_{L/R}} & =c_{L/R}
\begin{pmatrix}
0 \\ 1
\end{pmatrix},&
\tilde{\amp{u}}_{1/2_{L/R}} & =c_{L/R}^{\prime}
\begin{pmatrix}
1 \\ 0
\end{pmatrix},\label{tilde_u}\\
\tilde{\amp{v}}_{1/2_{L/R}} & =d_{L/R}
\begin{pmatrix}
0 \\ 1
\end{pmatrix},&
\tilde{\amp{v}}_{-1/2_{L/R}} & =d_{L/R}^{\prime}
\begin{pmatrix}
1 \\ 0
\end{pmatrix}.
\label{tilde_v}
\end{align}
In order to determine coefficients 
$c_{L/R}$, $c^\prime_{L/R}$, $d_{L/R}$, $d^\prime_{L/R}$,
we must use 
a manifestly covariant first order differential equation connecting the left and right 
handed chiral components of the neutrino field. 
This equation must be consistent with the above field construction procedure.

\section{Dirac-like equation}
\label{sec:Dirac-like-eq}

Now, we introduce 
an analog of the Dirac equation in our formalism
to conclude defining the free dynamics of the tachyonic neutrino field. 
A variety of the Lorentz covariant Dirac-like equations for spin-$\tfrac{1}{2}$ tachyons 
was discussed in Ref.~\cite{cab_Rembielinski1997}, however, 
only two equations satisfy the CPT invariance. 
Here we adopt the simplest of them, namely
\begin{equation}
(\gamma^5 \gamma^\mu i \partial_\mu - \kappa)\nu(x,u) = 0.
\label{Dirac_eq_neutrino_field}
\end{equation}
This equation was first introduced by Tanaka \cite{Tanaka_1960} while 
the corresponding Lagrangian was proposed by Chodos et al. \cite{ChHK1985_neutrino-tachyon}.

Equation (\ref{Dirac_eq_neutrino_field}) implies the space-like dispersion relation 
for $k_\mu$ and 
fixes interrelations between left and right components of the neutrino field. 
We use this equation for the  free field $\nu(x,u)$ introduced in Eq.~(\ref{nu_x_u_Fourier}) 
in the context of our approach to determine the coefficients 
$c_{L/R}$, $c_{L/R}^\prime$, $d_{L/R}$, $d_{L/R}^\prime$ 
of field amplitudes in Eqs.~(\ref{tilde_u},\ref{tilde_v}). 
Thus we obtain the following equations for the amplitudes
\begin{align}
&(\gamma^5 \gamma^\mu k_\mu - \kappa)\amp{u}_\lambda(k,u) = 0,
\label{Dirac_eq_u}\\
&(\gamma^5 \gamma^\mu k_\mu + \kappa)\amp{v}_\lambda(k,u) = 0,
\label{Dirac_eq_v}
\end{align}
where $k_\mu$ satisfies the covariant conditions (\ref{dispersion_tachyon}) and
(\ref{q_greater_zero}).	

For the chiral amplitudes, the above equations imply
\begin{align}
\amp{u}_{\lambda L}(k,u) & = - \tfrac{1}{\kappa} (k^0 I - 
\vec{k}\cdot \vecg{\sigma}) \amp{u}_{\lambda R}(k,u),
\label{ampl_u}\\
\amp{v}_{\lambda L}(k,u) & = \tfrac{1}{\kappa} (k^0 I - 
\vec{k}\cdot \vecg{\sigma}) \amp{v}_{\lambda R}(k,u),
\label{ampl_v}
\end{align}
and analogously for the space inverted pair.
These equations allow to determine the values of the coefficients
$c_{L/R}$, $c_{L/R}^\prime$, $d_{L/R}$, $d_{L/R}^\prime$ 
in Eqs.~(\ref{tilde_u},\ref{tilde_v}) for $k=\tilde{k}$ and $u=u_{PF}$. 
Taking into account (\ref{amp_u_generacja}) and (\ref{amp_v_generacja}),
we obtain the final form of the normalized amplitudes in an arbitrary frame $\Sigma_u$:
\begin{equation}
\amp{u}_{1/2}(k,u) = 
\tfrac{1}{\sqrt{2}}
\begin{pmatrix}
\sqrt{-q+\sqrt{q^2+\kappa^2}}\,\, 
\mathbb{L}_u \mathbb{U}_{\vec{n}}
\begin{pmatrix}
1 \\ 0
\end{pmatrix}\\
\sqrt{q+\sqrt{q^2+\kappa^2}}\,\, 
\mathbb{L}_{u^{\pi}} \mathbb{U}_{\vec{n}}
\begin{pmatrix}
1 \\ 0
\end{pmatrix}
\end{pmatrix},
\label{ampl_u_p_explicit}
\end{equation}
\begin{equation}
\amp{u}_{-1/2}(k,u) = 
\tfrac{1}{\sqrt{2}}
\begin{pmatrix}
\sqrt{q+\sqrt{q^2+\kappa^2}}\,\, 
\mathbb{L}_u \mathbb{U}_{\vec{n}}
\begin{pmatrix}
0 \\ 1
\end{pmatrix}\\
-\sqrt{-q+\sqrt{q^2+\kappa^2}}\,\, 
\mathbb{L}_{u^{\pi}} \mathbb{U}_{\vec{n}}
\begin{pmatrix}
0 \\ 1
\end{pmatrix}
\end{pmatrix},
\label{ampl_u_m_explicit}
\end{equation}
\begin{equation}
\amp{v}_{1/2}(k,u) = 
\tfrac{1}{\sqrt{2}}
\begin{pmatrix}
\sqrt{q+\sqrt{q^2+\kappa^2}}\,\, 
\mathbb{L}_u \mathbb{U}_{\vec{n}}
\begin{pmatrix}
0 \\ 1
\end{pmatrix}\\
\sqrt{-q+\sqrt{q^2+\kappa^2}}\,\, 
\mathbb{L}_{u^{\pi}} \mathbb{U}_{\vec{n}}
\begin{pmatrix}
0 \\ 1
\end{pmatrix}
\end{pmatrix},
\label{ampl_v_p_explicit}
\end{equation}
\begin{equation}
\amp{v}_{-1/2}(k,u) = 
\tfrac{1}{\sqrt{2}}
\begin{pmatrix}
\sqrt{-q+\sqrt{q^2+\kappa^2}}\,\, 
\mathbb{L}_u \mathbb{U}_{\vec{n}}
\begin{pmatrix}
1 \\ 0
\end{pmatrix}\\
-\sqrt{q+\sqrt{q^2+\kappa^2}}\,\, 
\mathbb{L}_{u^{\pi}} \mathbb{U}_{\vec{n}}
\begin{pmatrix}
1 \\ 0
\end{pmatrix}
\end{pmatrix},
\label{ampl_v_m_explicit}
\end{equation}
where $\mathbb{L}_u$ and $\mathbb{U}_{\vec{n}}$ are given
by Eqs.~(\ref{Ln_explicit}) and (\ref{Un_explicit}), respectively,
while $q$ and $\vec{n}$ are functions of $k$ and $u$, defined in 
Eq.~(\ref{q_def}) and Eq.~(\ref{vec_n_def}).

Using (\ref{ampl_u},\ref{ampl_v}) and the
appropriate formulas found in \ref{app:rotations}
one can easily check by means of
Eqs.~(\ref{dispersion_tachyon},\ref{u_square},\ref{q_def},\ref{q_greater_zero}), 
that the amplitudes (\ref{ampl_u_p_explicit},\ref{ampl_u_m_explicit},%
\ref{ampl_v_p_explicit},\ref{ampl_v_m_explicit}) fulfill 
Eqs.~(\ref{Dirac_eq_u},\ref{Dirac_eq_v}), i.e., the field $\nu(x,u)$ 
satisfies Eq.~(\ref{Dirac_eq_neutrino_field}). 
The scalar products of amplitudes and the polarization operators are collected in 
\ref{app:amplitudes}.

Dependence of the amplitudes
(\ref{ampl_u_p_explicit},\ref{ampl_u_m_explicit},%
\ref{ampl_v_p_explicit},\ref{ampl_v_m_explicit})
on $q$ is such that for $q\gg \kappa$ the following chiral
amplitudes vanish to high accuracy for each $u_\mu$
\begin{align}
&\amp{u}_{(1/2)L}(k,u) \to 0, && \amp{u}_{(-1/2)R}(k,u) \to 0,
\label{u_pL_zero}\\
&\amp{v}_{(1/2)R}(k,u) \to 0, && \amp{v}_{(-1/2)L}(k,u) \to 0,
\label{v_pR_zero}
\end{align}
from which the following limiting forms are obtained for $q\gg\kappa$
\begin{align}
&\amp{u}_{-1/2} \to \tfrac{1}{2}(1-\gamma^5) \amp{u}_{-1/2},
&&\amp{v}_{1/2} \to \tfrac{1}{2}(1-\gamma^5) \amp{v}_{1/2},
\label{u_m_limit}\\
&\amp{u}_{1/2} \to \tfrac{1}{2}(1+\gamma^5) \amp{u}_{1/2},
&&\amp{v}_{-1/2} \to \tfrac{1}{2}(1+\gamma^5) \amp{v}_{-1/2}.
\label{u_p_limit}
\end{align}
Thus, anticipating the Standard Model dynamics where only the  left-handed chirality 
of the neutrino and the right-handed chirality of the antineutrino participate, 
we conclude that one can effectively observe
exactly only  neutrinos with helicity $-\tfrac{1}{2}$ and antineutrinos with helicity $\tfrac{1}{2}$.
If the mass of a space-like neutrino is about or less than
1~eV  then this condition is fulfilled already for
$q\sim$ tens of eV so we reproduce 
the property observed experimentally in the MeV and GeV energy range since ever.

On the other hand, if $q \sim \kappa$ then 
the dependence of the amplitudes (\ref{ampl_u_m_explicit},\ref{ampl_u_p_explicit},%
\ref{ampl_v_m_explicit},\ref{ampl_v_p_explicit}) on $q$ indicates that
both neutrino
helicities are present and in consequence one can expect specific predictions to 
be different from those based on the SM. 
This  regards in particular the electron energy spectrum near the endpoint in 
$\beta$ decay with the tachyonic neutrino (see Sec.~\ref{sec:beta-decay}).
The $q$-dependence of the tachyonic neutrino amplitudes was discussed 
firstly in Ref.~\cite{cab_Rembielinski1997} in the context of the reduced model 
with only one helicity component ($-\tfrac{1}{2}$ for neutrino and 
$\tfrac{1}{2}$ for antineutrino, respectively).

\section{Helicity observable}
\label{sec:helicity}

The helicity operator $\Hat{\lambda}$ given in Eq.~(\ref{lambda_def})
is realized in the bispinor space as $\Hat{\Lambda}\sim u_\mu W^\mu$, where
$W^\mu = -\tfrac{1}{2} \varepsilon^{\mu\nu\tau\sigma} 
P_\nu S_{\tau\sigma}$ is the Pauli--Lubanski  pseudo-vector. 
In the coordinate representation it reads
\begin{equation}
\Hat{\Lambda}(u) \sim \gamma^5 [\gamma^\nu (i\partial_\nu),u_\mu \gamma^\mu].
\end{equation}
After normalizing in the momentum representation it takes the form
\begin{align}
\Hat{\Lambda}(k,u) & = \frac{\mp1}{4\sqrt{q^2+\kappa^2}}
\gamma^5 [k_\mu \gamma^\mu,u_\nu \gamma^\nu]
\label{Lambda_gamma_1}\\
&= \mp S(L_u R_{\vec{n}}) \frac{1}{2}
\begin{pmatrix}
\sigma^3 & 0 \\ 0 & \sigma^3
\end{pmatrix}
\overline{S(L_u R_{\vec{n}})},
\label{Lambda_gamma_2}
\end{align}
where the sign $-$ or $+$ is related to the negative ($\amp{v}$) or 
positive ($\amp{u}$)
frequencies in Eq.~(\ref{nu_x_u_Fourier}). 
In the PF the Weyl representation  of $\Hat{\Lambda}$ has 
a simple form
\begin{equation}
\Hat{\Lambda}(k,u_{PF}) = \mp \tfrac{1}{2}
\begin{pmatrix}
\tfrac{\vec{k}}{|\vec{k}|}\cdot \vecg{\sigma} & 0\\
0 & \tfrac{\vec{k}}{|\vec{k}|}\cdot \vecg{\sigma}
\end{pmatrix}.
\end{equation}
The action of $\Hat{\Lambda}(u,k)$ on the amplitudes can be obtained from 
(\ref{Lambda_gamma_1},\ref{Lambda_gamma_2}) and
(\ref{ampl_u_p_explicit},\ref{ampl_u_m_explicit},%
\ref{ampl_v_p_explicit},\ref{ampl_v_m_explicit}).

\section{Discrete symmetries}
\label{sec:discrete-sym}

It is easy to see that the standard space inversion of a bispinor  
\begin{equation}
\nu_P(x,u) = \gamma^0 \nu(x^\pi,u^\pi),
\label{parity_standard}
\end{equation}
where $x^\pi=(x^0,\vec{x})^\pi=(x^0,-\vec{x})$ and similarly for other four-vectors,
does not preserve the form of 
the Dirac-like equation (\ref{Dirac_eq_neutrino_field});
instead, this equation is form-invariant with respect to the parity transformation
defined as 
\begin{equation}
\nu_P(x,u) = i \gamma^5 \gamma^0 \nu(x^\pi,u^\pi).
\label{parity}
\end{equation}
The Lagrangian
\begin{equation}
\mathcal{L} = \overline{\nu(x,u)} 
\big(
\gamma^5 (\gamma^\mu i \partial_\mu) - \kappa
\big)
\nu(x,u)
\end{equation}
related to Eq.~(\ref{Dirac_eq_neutrino_field}) 
is not invariant under the standard inversion (\ref{parity_standard}) and changes sign 
under the parity transformation (\ref{parity}) so both operations 
lead to parity nonconservation.

Now, if we introduce the charge conjugation according to the standard procedure
(see also Ref.~\cite{ChHK1985_neutrino-tachyon}) as
\begin{equation}
\nu_C(x,u) = \mathcal{C} \overline{\nu(x,u)^T},
\label{nu_C}
\end{equation}
where the unitary matrix $\mathcal{C}$ satisfies the relation 
$\mathcal{C} \gamma^{\mu T} \mathcal{C}^T = \gamma^\mu$ 
in this case
(in the Weyl representation $\mathcal{C} = i \gamma^5 \gamma^2 \gamma^0 
= \gamma^3 \gamma^1$), the Lagrangian $\mathcal{L}$ is also non invariant. 
Moreover, neither the parity nor charge conjugation operations can be realized 
as symmetries of the tachyonic neutrino even on the free level.

Fortunately, it is possible to define a combination of the parity and charge conjugation, CP
\begin{equation}
\nu_{CP}(x,u) = \gamma^5 \gamma^0 \nu_C (x^\pi,u^\pi).
\end{equation}
Thus
\begin{multline}
\nu_{CP}(x,u) = \tfrac{1}{(2\pi)^{3/2}} \int d\mu(k,u)
\sum_\lambda 
\big[
e^{ikx} \amp{v}_\lambda(k,u) \times\\
a^\dagger_\lambda(k^\pi,u^\pi)
+ e^{-ikx} \amp{u}_\lambda(k,u) 
b_\lambda(k^\pi,u^\pi)
\big],
\label{nu_CP_Fourier}
\end{multline}
where we have used the relations  
$\amp{u}_\lambda (k^\pi,u^\pi ) = \gamma^5 \gamma^0 \amp{u}_\lambda (k,u)$
and  
$\amp{v}_\lambda (k^\pi,u^\pi ) = - \gamma^5 \gamma^0 \amp{v}_\lambda (k,u)$
(see \ref{app:amplitudes}).

Next, we connect $\nu_{CP} (x,u)$ with the action of the unitary operator 
$\mathcal{Q}$ representing the CP transformation
\begin{equation}
\zeta \nu_{CP}(x,u) \equiv \mathcal{Q} \nu(x,u) \mathcal{Q}^\dagger,
\end{equation}
where $\zeta$ is  a phase factor. Consequently, by means of Eq.~(\ref{nu_CP_Fourier})
\begin{align}
&\mathcal{Q} a_\lambda(k,u) \mathcal{Q}^\dagger 
= \zeta b_\lambda(k^\pi,u^\pi), \\
&\mathcal{Q} b_\lambda(k,u) \mathcal{Q}^\dagger 
= \zeta^* a_\lambda(k^\pi,u^\pi)
\end{align}
and the remaining relations are obtained by the Hermitian conjugation and exchange 
$(k,u)$ with $(k^\pi,u^\pi)$. 
Concluding, in the case of a  tachyonic neutrino one cannot define
separate discrete symmetries C or P; but only the CP symmetry can be realized.

\section{Interactions}
\label{sec:interactions}

Let us start with an observation made in the paper by 
Chodos \textit{et al.}\ \cite{ChHK1985_neutrino-tachyon} that 
the Higgs mechanism can, 
in principle, lead to a tachyonic fermion as well as to a massive fermion. 
Indeed, the kinetic  part of the massless bispinor field Lagrangian could have two 
different forms
\begin{equation}
\bar{\psi} (\gamma^\mu i \partial_\mu) \psi \quad \text{or}\quad
\bar{\psi} \gamma^5 (\gamma^\mu i \partial_\mu) \psi
\end{equation}
leading to equivalent massless Dirac equations  
\begin{equation}
(\gamma^\mu i \partial_\mu) \psi = 0 \quad \text{and}\quad
\gamma^5 (\gamma^\mu i \partial_\mu) \psi = 0.
\end{equation}
However, mass generation \textit{via} the Yukawa coupling with the Higgs fields 
leads to inequivalent free field Lagrangians
\begin{equation}
\bar{\psi} (\gamma^\mu i \partial_\mu) \psi - m \bar{\psi} \psi
\quad \text{and}\quad
\bar{\psi} \gamma^5 (\gamma^\mu i \partial_\mu) \psi - \kappa \bar{\psi} \psi,
\end{equation}
respectively. 
The first Lagrangian results in the standard Dirac equation while the second 
Lagrangian leads to the free field equation 
(\ref{Dirac_eq_neutrino_field}) describing a tachyonic fermion. 

On the other hand as was pointed out  in
\cite{CRS2006_Decays_spacelike_neutrinos},
before the onset of the spontaneous symmetry breaking process, 
massless  Lagrangians for the neutrino field $\nu$ and the charged lepton field $l$ 
had the form $\mathcal{L}_0^I$ in the standard case
(leading to two Dirac leptons, $l$ and $\nu$)
and $\mathcal{L}_0^{II}$ in the mixed case 
(leading to a Dirac lepton and a tachyonic neutrino)
\begin{align}
& \mathcal{L}_0^{I} = \bar{l} (\gamma^\mu i \partial_\mu) l 
+ \bar{\nu} (\gamma^\mu i \partial_\mu) \nu,\\
& \mathcal{L}_0^{II} =\bar{l} (\gamma^\mu i \partial_\mu) l 
+ \bar{\nu} \gamma^5 (\gamma^\mu i \partial_\mu) \nu.
\label{Lagrangian_L_0_II}
\end{align}
Now, in the chiral representation of the fields $\nu$ and $l$,
Lagrangian $\mathcal{L}_0^{I}$ is invariant under
$(\sg{SU}(2)_L \times \sg{U}(1)_L)\times (\sg{SU}(2)_R \times \sg{U}(1)_R)$
transformations  of the chiral left doublet $L$ and the right doublet $R$. 
On the other hand, Lagrangian
$\mathcal{L}_0^{II}$
is invariant under
$(\sg{SU}(2)_L \times \sg{U}(1)_L)$
transformation of the left doublet $L$ and under
$(\sg{U}(1)_{I_3 R} \times \sg{U}(1)_R)$
subgroup of the right
$(\sg{SU}(2)_R \times \sg{U}(1)_R)$
group. 
Therefore, $\mathcal{L}_0^{II}$ admits only a doublet $L$ and two singlets,
$\nu_R$ and $l_R$, exactly as one needs for a formulation of the electroweak 
sector of the Standard Model. 
Concluding, the mixed Lagrangian (\ref{Lagrangian_L_0_II}) 
fixes exactly the weak group and its 
representation without additional requirements.

By means of the standard procedure of gauging and the spontaneous symmetry 
breaking up to the electromagnetic gauge group $\sg{U}(1)_{QED}$ we obtain 
a Lagrangian which differs from the SM Lagrangian by the neutrino kinetic term only. 
In the unitary gauge the leptonic part of this Lagrangian takes the usual form
\begin{multline}
\mathcal{L}_\sg{lepton} = 
\bar{l}(\gamma^\mu i \partial_\mu - m_l)l
+
\bar{\nu}(\gamma^5 \gamma^\mu i \partial_\mu - \kappa)\nu\\
+ \tfrac{g}{2\sqrt{2}} (W_{\mu}^- j_{+}^\mu + 
W_{\mu}^+ j_{-}^\mu)
+ \tfrac{g}{2 \cos \theta_W} Z_\mu (j_{l}^\mu + j_{\nu L}^\mu)\\
- e A_\mu j_{e}^\mu
- \tfrac{m_l}{v} \bar{l} H l
- \tfrac{\kappa}{v} \bar{\nu} H \nu,
\label{Lagrangian-lepton}
\end{multline}
where $g=e/\sin\theta_W$, $v^2={1}/{(\sqrt{2}G_F)}$,
$e$ is the electric charge, 
$G_F$ -- the Fermi constant,
$\theta_W$ -- the Weinberg angle, 
$m_l$ -- mass of the lepton $l$.
Here, $H$ denotes the Higgs field, 
$W_{\mu}^\pm$ and $Z_\mu$ are the charged and neutral 
weak bosons fields, respectively, 
and $A_\mu$ the electromagnetic four-potential. 
The electroweak currents take also the standard form
$j_{+}^\mu  = \bar{l} \gamma^\mu(1-\gamma^5)\nu$,
$j_{-}^\mu = \bar{\nu} \gamma^\mu(1-\gamma^5)l$,
$j_{e}^\mu = \bar{l} \gamma^\mu l$,
$j_{l}^\mu = \bar{l} \gamma^\mu (g_V - g_A \gamma^5) l$,
$j_{\nu L}^\mu = \tfrac{1}{2} \bar{\nu} \gamma^\mu
(1-\gamma^5) \nu$,
$j_{\nu R}^\mu = \tfrac{1}{2} \bar{\nu} \gamma^\mu
(1+\gamma^5) \nu$,
where $g_V=2\sin^2\theta_W-\tfrac{1}{2}$, $g_A=-\tfrac{1}{2}$.
Taking into account the form of the neutrino amplitudes discussed in 
Sec.~\ref{sec:Dirac-like-eq}, Lagrangian (\ref{Lagrangian-lepton}), 
at least in the tree approximation, leads to the SM results 
for energies significantly higher than the neutrino mass $\kappa$
(as well as to an agreement with the model reduced to the one helicity 
component discussed in 
\cite{cab_Rembielinski1997,CR1999_beta_decay_tachyonic_neutrino}). 
However, for energies close to the tachyonic neutrino mass $\kappa$, 
where both the left and right handed chiral components are present, 
one could expect new effects.

\section{Beta decay of ${}^3$H}
\label{sec:beta-decay}

Tritium decay with a space-like neutrino was discussed by us within the framework 
of the reduced model \cite{CR1999_beta_decay_tachyonic_neutrino};
the corresponding differential decay rate (electron energy spectrum) 
shows an anomalous  energy dependence close to  the endpoint. 
Earlier calculations of the Kurie plot
\cite{ChHK1985_neutrino-tachyon} also indicated 
differences compared to the prediction for a massive neutrino. 
Below we derive the corresponding expressions for the complete model, i.e., with 
the lepton interaction Lagrangian given by (\ref{Lagrangian-lepton}), 
taking into account the contributions of all two chiral and helicity components of 
the tachyonic neutrino field.

The amplitude squared of this process, $|M|^2$, is given at the tree level
by the formula
\begin{multline}
|M|^2 \sim G_{F}^2 
\tr\big[
(l\gamma+m_e)\gamma^\mu(1-\gamma^5)
(\gamma^5 k\gamma + \kappa)\times\\
\gamma^\nu (1-\gamma^5)
\big]
\tr\big[
(p\gamma+m_P)\gamma_\mu (I-g_A \gamma^5) \times\\
(r\gamma+m_N) 
\gamma_\nu(I-g_A \gamma^5)
\big],
\end{multline}
i.e.,
\begin{multline}
|M|^2 \sim 2 g_A [(lr)(kp)-(lp)(kr)] 
-(1+g_A^2)[(lp)(kr)\\
+(lr)(kp)]
+(1-g_A^2)m_N m_P (kl),
\end{multline}
where $m_N$, $m_P$, $m_e$, $\kappa$ and $r$, $p$, $l$, $k$ are the masses and
four-momenta of tritium ${}^3$H, helium ${}^3$He, electron and tachyonic 
anti-neutrino, respectively.
By means of the above formulas one can obtain the differential decay rate in terms of the 
outgoing electron energy, $E$, in the form
\begin{equation}
\dfrac{d\Gamma}{dE} = 
\dfrac{1}{128 \pi^3 m_N}
\int_{\max\{0,k_-\}}^{k_+}
|M|^2 \,dq,
\end{equation}
where, the limiting values of the neutrino 
energy following from the tachyonic kinematics in the preferred frame are 
%\begin{widetext}
\begin{multline}
	k_\pm = 
	\Big[(m_N-l_0)(\Delta m^2 - 2 l^0 m_N)
		\pm 
		\Big\{({l^0}^2-m_{e}^2)\times\\
		    \big[
			(\Delta m^2 - 2 l^0 m_N)^2
			+ 4 \kappa^2 (m_{e}^2 + m_{N}^2 - 2 l^0 m_N)	
			\big]\Big\}\Big]\times\\
	\Big[2(m_{e}^2+m_{N}^2-2 l^0 m_N)\Big]^{-1},
\end{multline}
%\end{widetext}%
where $l^0=E+m_e$ and $\Delta m^2 = (m_{N}^2-m_{P}^2)
+(m_{e}^2-\kappa^2)$.

\begin{figure}
\begin{center}
\includegraphics[width=1\columnwidth]{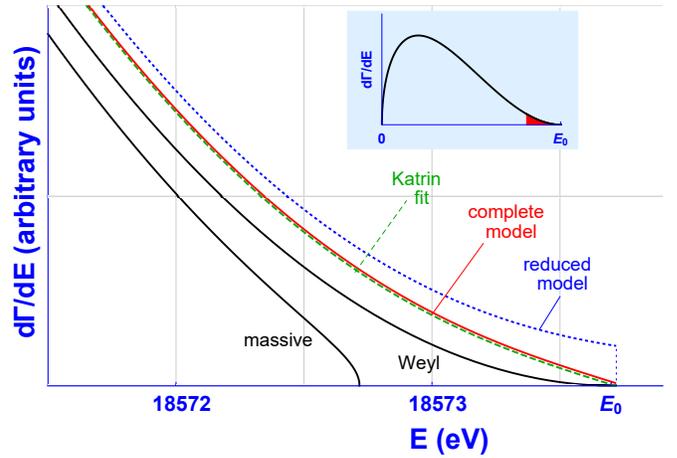}
\end{center}
\caption{Theoretical electron energy spectrum (differential decay rate)
	near the endpoint in tritium decay at rest;
	$E_0=18573.7$~eV is the endpoint energy fitted by KATRIN. The curves  demonstrate  the cases for: massive   neutrino  with $m_{\nu}=1$~eV
	(the kinematical limit in  the decay with a massive neutrino is $E_0-m_{\nu}$); 
	massless (Weyl)
	neutrino; massive neutrino  with inverted sign of  $m_{\nu}^2$ illustrating the outcome of the KATRIN fit to their data (dashed line); complete model (solid line)
	and the reduced model (dotted line) for a tachyonic neutrino with $\kappa=1$~eV. 
	The curves for the complete model and the KATRIN fit overlap 
	in reality but have been infinitesimally shifted  for the purpose of illustration.
    Inset: figurative  visualization of the range covered on the main plot 
    w.r.t. the full electron energy spectrum within bounds [$0,E_0$].}
\label{Fig1}
\end{figure}

The above differential decay rate $d\Gamma/dE$,
as a function of the electron kinetic energy,
is presented in Fig.~\ref{Fig1} 
together with the corresponding curve predicted in the reduced
model \cite{CR1999_beta_decay_tachyonic_neutrino}. 
We also  show curves for a massive neutrino and 
massless Weyl neutrino. 
The predictions of the complete and the reduced models are 
slightly different near the endpoint
which allows (under a working hypothesis that the neutrino is a
tachyonic fermion) to test both possibilities experimentally. 

Recently the KATRIN Collaboration delivered the most precise measurement of the 
neutrino mass squared in tritium decay: 
$m_{\nu_e}^2 = -1.0_{-1.1}^{+0.9}$~eV${}^2$
\cite{Aker_KATRIN2019_PhysRevLett.123.221802,aker2019_KATRIN_tritium_first}.
The prediction of the complete model  with $\kappa^2=1$~eV$^2$ and the 
recent  KATRIN fit representing the above value are indistinguishable. 
The KATRIN result, despite its yet insufficient statistical significance, 
is a subsequent one in a row yielding negative central values for the fitted neutrino 
mass squared, 
following earlier measurements at Mainz \cite{Weinheimer_etal1999,Kraus_etal2005_Mainz} 
and Troitsk \cite{Lobashev_etal1999_Troitsk,Aseev_etal2011_Troitsk_PhysRevD.84.112003}.

\section{Conclusions}

We have presented for the first time a fully consistent formalism of quantization 
of space-like fermions with helicity $\pm\tfrac{1}{2}$. 
In our formalism all common concerns have been resolved and explained.
Specifically, (i) negative energy problem is solved and no imaginary energies appear, 
(ii) causality paradoxes do not appear, 
(iii) the proper quantization procedure has been described in detail 
(Hilbert space with suitable  Poincare properties indicated), 
(iv) interactions are unitary and theory is perturbatively renormalizable (anomaly free), 
(v) since the presented description is Lorentz covariant, all properties apply 
to arbitrary reference frames (and not only the preferred frame).
In order to achieve that and to avoid causal problems with tachyons, 
existence of a preferred frame must 
be assumed; it is natural to suppose that the PF coincides with
the Cosmic Neutrino Background  frame. 
We again indicate that the preferred frame, if exists in nature, would be uniquely 
a part of the tachyonic sector and not that of conventional physics that 
describes light and slower than light (massive) objects.
For those  the preferred frame is but an ordinary inertial frame respecting 
the principles of the Einsteinian relativity. 
The preferred frame cannot be discovered with the use of light or 
massive particles owing to the intrinsic properties of the Einsteinian relativity, 
in particular the corresponding clock synchronisation procedure.
We determined the unitary realizations of the inhomogenous Lorentz group by means 
of the Wigner-Mackey induction procedure and constructed the corresponding Fock space. 
In the preferred frame scenario, the irreducible unitary realizations of the 
Lorentz group for tachyons are labeled by particle helicity (not by spin). 
The ultimate, manifestly Lorentz covariant formalism was developed by way 
of constructing  the field amplitudes according to the Weinberg method. 
Since our theoretical findings regarding the space-like helicity 
$\pm\tfrac{1}{2}$ fermions suggest 
connotations with neutrinos, we show that indeed the following facts and observations 
from the field of neutrino physics can be interpreted within the presented approach 
at the level of fundamental properties. 
According to our formalism, only neutrinos with helicity $-\tfrac{1}{2}$  and antineutrinos 
with helicity $+\tfrac{1}{2}$ can be effectively observed at asymptotic energies 
(significantly exceeding the neutrino mass, $\kappa$, amounting to or 
a fraction of a single eV) which is a well established knowledge based on the lack 
of experimental evidence for the complementary cases at MeV--and higher energies. 
It should be stressed that in our formalism this fact would be a consequence solely 
of the ``tachyoneity'' of the neutrino, i.e., its intrinsic property at the free level and 
not exclusively due to the observed character of weak interactions. 
Likewise would be the chiral asymmetry of weak interactions, thus far introduced 
in an explicit way to the structure of the neutrino current. 
These inherent features of space-like fermions follow directly from the energy 
dependence of the amplitudes for asymptotic energies $q\gg\kappa$, 
described by Eqs.~(\ref{u_pL_zero},\ref{v_pR_zero},\ref{u_m_limit},\ref{u_p_limit}).
This allows to explain the
mysterious chiral properties of the neutrinos as well as occurrence of only 
$-\tfrac{1}{2}$ helicity neutrino and its conjugate in nature, 
both age-old experimental observations. 
In addition, we have also shown that neither C nor P symmetry holds for 
space-like helicity $\pm\tfrac{1}{2}$ fermions at the free level but rather the combined 
CP symmetry is conserved. 
The nature of this property is intrinsic if neutrinos are tachyons and not arising 
due to their interactions.

As was
mentioned in Sec.~\ref{sec:interactions}, the mixed free field 
Lagrangian (\ref{Lagrangian_L_0_II}) yet before of the spontaneous symmetry 
breaking admits only the Weinberg--Salam weak group and its representation 
without additional  requirements.
After gauging and the electroweak symmetry breaking in the lepton sector 
one obtains finally the Lagrangian (\ref{Lagrangian-lepton}), 
identical to the SM Lagrangian except of the neutrino kinetic term. 
The resulting model is anomaly free and perturbatively renormalizable.

Tachyonic neutrinos also can oscillate according to the same pattern as massive 
neutrinos; oscillations in the present formulation  do not
distinguish between the massive and tachyonic  forms of the dispersion
rela\-tion~\cite{CRSW_2006_Oscillations}.

In the low energy neutrino limit, the negative value of the neutrino mass squared, 
determined in several recent tritium decay experiments, makes an interesting 
observation, though not yet conclusive but pointing to the hypothesis of 
a space-like nature of the neutrino.
As we have shown in Fig.~\ref{Fig1}, the conventional fit, recently confirmed by KATRIN,
coincides with the prediction of our complete model 
under the assumption that the tachyonic neutrino mass amounts to $\kappa=1$~eV. 

On the other hand, tachyonic neutrinos are the Dirac particles in our approach. 
Thus one way of falsifying the hypothesis that neutrinos are tachyons would 
be to observe the neutrinoless double beta decay indicating that neutrinos 
were the Majorana particles. 
This process has not been discovered to date despite several decades of experimenting.

\appendix

\section{Invariant measure, the spherical Dirac delta and completeness relation}
\label{app:measure}

The Lorentz invariant measure taking into account existence of the preferred 
frame has the following form
\begin{align}
d\mu(k,u) & = d^4k\, \delta(k^2+\kappa^2) \Theta(uk)
\nonumber\\
& = dk^0 |\vec{k}|^2\, d|\vec{k}|\, d\Omega\, 
\Theta(uk) \delta(|\vec{k}|^2 - \omega^2) 
\nonumber\\
& = dk^0 |\vec{k}|^2\, d|\vec{k}|\, d\Omega\, 
\Theta(uk) \dfrac{\delta(|\vec{k}|-\omega)}{2\omega}
\nonumber\\
& = \dfrac{\omega(q,u,\theta)}{2 u^0} \Theta(q) \, dq\, d\Omega.
\end{align}
Here $\theta$ is the angle between $\vec{u}$ and $\vec{k}$, 
$\omega(q,u,\theta)$ is given in Eq.~(\ref{omega_k_u}). 
Furthermore $\Theta$ is the Heaviside Theta and
\begin{align}
d^4 k & = d k^0 \wedge d^3 \vec{k} \nonumber\\
 & = \tfrac{1}{u^0} d(q+\vec{u}\cdot\vec{k}) \wedge d^3 \vec{k} \nonumber \\
 & = \tfrac{1}{u^0} dq \wedge d^3 \vec{k} \nonumber \\
 & = \tfrac{1}{u^0} dq d^3 \vec{k}.
\end{align}
The spherical Dirac delta is the angular part of the three dimensional 
Dirac delta. It is defined as follows: 
Let $\vec{n}(\theta,\varphi)$
denote a unit vector depending on the spherical coordinates (angles) 
$\theta$ and $\varphi$. 
Then the spherical Dirac delta is defined by the formula
\begin{equation}
\int_{S_2} d\Omega\, \delta(\vec{n}-\vec{n}_0) f(\vec{n}) 
= f(\vec{n}_0), 
\label{Dirac-delta-spherical}
\end{equation}
where
$\vec{n}_0 = \vec{n}(\theta_0,\varphi_0)$
and $d\Omega$ is the solid angle differential. 
In terms of the angles $\theta$ and $\varphi$
$d\Omega = \sin\theta\,d\theta\,d\varphi$
and
$\delta(\vec{n}-\vec{n}_0)=
\tfrac{\delta(\theta-\theta_0)\delta(\varphi-\varphi_0)}{\sin\theta}$.

Below we prove explicitly the completeness relation (\ref{completeness-relation}).
Since neutrino and antineutrino spaces are mutually orthogonal, it is enough to prove
this relation in one of those spaces, say neutrino space of states.
Therefore, we are to prove that for any vector 
\begin{equation}
\ket{\psi,u} = \sum_{\lambda^\prime} \int d\mu(p,u) \psi(p,u,\lambda^\prime)
\ket{p,u,\lambda^\prime} 
\end{equation}
from one-particle neutrino space
it holds
\begin{equation}
\sum_\lambda \int d\mu(k,u) 
\Big(\ket{k,u,\lambda}\bra{k,u,\lambda} \Big) \ket{\psi,u}
= \ket{\psi,u}.
\label{to-prove}
\end{equation}
Indeed, we have
\begin{align}
&(\ket{k,u,\lambda}\bra{k,u,\lambda})\ket{\psi,u} \nonumber\\
& = \sum_{\lambda^\prime} \int d\mu(p,u) \psi(p,u,\lambda^\prime)
\bracket{k,u,\lambda}{p,u,\lambda^\prime} \ket{k,u,\lambda}\nonumber\\
& = \sum_{\lambda^\prime} \int dq_p\, d\Omega(\theta_p,\varphi_p)\,\,
\frac{\omega(q_p,u,\theta_p)}{2u^0} \Theta(q_p)  \times\nonumber\\
&\phantom{=\sum_{\lambda^\prime} \int} 
\frac{2u^0}{\omega(q_k,u,\theta_k)}
\delta_{\lambda\lambda^\prime} \delta(q_k-q_p) 
\delta(\Hat{\vec{k}}-\Hat{\vec{p}})\times\nonumber\\
&\phantom{=\sum_{\lambda^\prime} \int \frac{2u^0}{\omega(q_k,u,\theta_k)}
	\delta_{\lambda\lambda^\prime}} 
\psi(p,u,\lambda^\prime) \ket{k,u,\lambda}\nonumber\\
& = \psi(k,u,\lambda) \ket{k,u,\lambda},
\end{align}
where we used Eqs.~(\ref{measure_inv},%
\ref{normalization},\ref{Dirac-delta-spherical}).
This last equation immediately implies Eq.~(\ref{to-prove}).

\section{Rotations, boosts etc.}
\label{app:rotations}

The rotation $R_{\vec{n}}$ reduced to the space sector has the form 
\begin{equation}
R_{\vec{n}} = 
\begin{pmatrix}
1-\tfrac{(n^1)^2}{1+n^3} & -\tfrac{n^1 n^2}{1+n^3} & n^1\\
-\tfrac{n^1 n^2}{1+n^3} & 1-\tfrac{(n^2)^2}{1+n^3} & n^2 \\
-n^1 & -n^2 & n^3
\end{pmatrix}
\end{equation}
while its spinor counterpart $\mathbb{U}_{\vec{n}}$
\begin{equation}
\mathbb{U}_{\vec{n}} = \dfrac{1}{\sqrt{2(1+n^3)}}
\begin{pmatrix}
1+n^3 & -n^1+in^2\\
n^1+in^2 & 1+n^3
\end{pmatrix}.
\end{equation}
Here the unit vector $\vec{n}(k,u)$ is given by Eq.~(\ref{vec_n_def}).
Boost and its spinor representative have the form
\begin{align}
L_u & = 
\begin{pmatrix}
u^0 & \vec{u}^T \\
\vec{u} & I+\frac{\vec{u}\otimes\vec{u}^T}{1+u^0}
\end{pmatrix},\\
\mathbb{L}_u & =
\dfrac{1}{\sqrt{2(1+u^0)}}
\Big(
(1+u^0)I - \vec{u}\cdot\vecg{\sigma}
\Big).
\end{align}
One can also show that the following relations hold
\begin{gather}
\mathbb{L}_u^\dagger = \mathbb{L}_u,\quad
\mathbb{L}_{u^\pi} = \mathbb{L}_{u}^{-1}\\
S(L_u R_{\vec{n}}) = 
\begin{pmatrix}
\mathbb{L}_u \mathbb{U}_{\vec{n}} & 0 \\
0 & \mathbb{L}_{u}^{-1} \mathbb{U}_{\vec{n}}
\end{pmatrix},\\
S^{-1}(L_u R_{\vec{n}}) = \overline{S(L_u R_{\vec{n}})}.
\end{gather}
Let
$k\sigma = k^0 I - \vec{k}\cdot\vecg{\sigma}$,
$k^\pi\sigma = k^0 I + \vec{k}\cdot\vecg{\sigma}$,
and
$\tilde{k}\sigma = qI - \sqrt{q^2+\kappa^2}\,\sigma^3$.
Then
\begin{align}
k\sigma & = \mathbb{L}_u \mathbb{U}_{\vec{n}} (\tilde{k}\sigma) 
\mathbb{U}_{\vec{n}}^\dagger \mathbb{L}_u\nonumber\\
& = \mathbb{L}_u (qI - \sqrt{q^2+\kappa^2}\,
\vec{n}\cdot\vecg{\sigma}) \mathbb{L}_u,\\
k^\pi\sigma & = \mathbb{L}_{u}^{-1} 
\mathbb{U}_{\vec{n}} (\tilde{k}^\pi\sigma) 
\mathbb{U}_{\vec{n}}^\dagger 
\mathbb{L}_{u}^{-1}\nonumber\\
& = \mathbb{L}_{u}^{-1} (qI + \sqrt{q^2+\kappa^2}\,
\vec{n}\cdot\vecg{\sigma}) \mathbb{L}_{u}^{-1},
\end{align}
and
\begin{equation}
\mathbb{L}_u \sigma^0 \mathbb{L}_u
= \mathbb{L}_u^2
=u^0 I - \vec{u}\cdot\vecg{\sigma},
\end{equation}
where
\begin{equation}
\vec{n}\cdot\vecg{\sigma} = 
\begin{pmatrix}
n^3 & n^1-in^2 \\ n^1+in^2 & -n^3
\end{pmatrix}
=\mathbb{U}_{\vec{n}} \sigma^3 \mathbb{U}_{\vec{n}}^\dagger
\end{equation}
and 
\begin{equation}
\vec{n} = \vec{n}(k,u) 
= \frac{1}{\sqrt{q^2+\kappa^2}} 
\left(
\vec{k} - \frac{q+k^0}{1+u^0}\vec{u}
\right).
\end{equation}
Moreover
\begin{equation}
\sigma^2 \mathbb{U}_{\vec{n}}^* \sigma^2 = \mathbb{U}_{\vec{n}},\qquad
\sigma^2 \mathbb{L}_{u}^T \sigma^2 = \mathbb{L}_{u}^{-1}.
\end{equation}
{}

\section{Relations between amplitudes: Space-inverted amplitudes, 
scalar products and polarization operators}
\label{app:amplitudes}

The amplitudes given in (\ref{ampl_u_p_explicit},\ref{ampl_u_m_explicit},%
\ref{ampl_v_p_explicit},\ref{ampl_v_m_explicit})
satisfy a number of relations listed in this Appendix. 
Taking into account the fact that in the case of the space inverted pair
$(k^\pi,u^\pi)$, we have
$(k^\pi,u^\pi)=\mathbb{L}_{u^\pi}R_{\vec{n}}(\tilde{k}^\pi,u_{PF})$,
where $\tilde{k}^\pi=(q,0,0,-\sqrt{q^2+\kappa^2})$,
we can obtain space inverted amplitudes. 
To do this, 
$\tilde{k}$ should be replaced by $\tilde{k}^\pi$ and $\mathbb{L}_u$ by
$\mathbb{L}_{u^\pi}$ in the formulas (\ref{amp_u_generacja},\ref{amp_v_generacja}). 
Notice, that the following relations hold
\begin{align}
\amp{u}_\lambda(k^\pi,u^\pi) & = \gamma^5 \gamma^0 \amp{u}_\lambda(k,u),\\
\amp{v}_\lambda(k^\pi,u^\pi) & = -\gamma^5 \gamma^0 \amp{v}_\lambda(k,u)
=-\gamma^0 \amp{u}_{-\lambda}(k,u),\\
\amp{u}_\lambda(k,u) & = - \gamma^5 \amp{v}_{-\lambda}(k,u).
\end{align}
The scalar products of the amplitudes take the form
\begin{align}
&\overline{\amp{u}_\lambda(k,u)} \amp{u}_\sigma(k,u)  = 2 \lambda \delta_{\lambda\sigma},\\
&\overline{\amp{v}_\lambda(k,u)} \amp{v}_\sigma(k,u)  = 2 \lambda \delta_{\lambda\sigma},\\
&\overline{\amp{u}_\lambda(k,u)} \amp{v}_\sigma(k,u) = 
\overline{\amp{v}_\lambda(k,u)} \amp{u}_\sigma(k,u) =0,\\
&\overline{\amp{u}_\lambda(k,u)} \gamma^5 \gamma^0 \amp{u}_\sigma(k,u) 
= - 2 \lambda k^0 \delta_{\lambda\sigma},\\
&\overline{\amp{v}_\lambda(k,u)} \gamma^5 \gamma^0 \amp{v}_\sigma(k,u) 
= 2 \lambda k^0 \delta_{\lambda\sigma},\\
&\overline{\amp{u}_\lambda(k,u)} \gamma^5 \gamma^0 \amp{v}_\sigma(k^\pi,u^\pi) 
=0,\\
&\overline{\amp{v}_\lambda(k,u)} \gamma^5 \gamma^0 \amp{u}_\sigma(k^\pi,u^\pi) 
=0.
\end{align}
\begin{multline}
\big(
\kappa + \gamma^5 (\gamma^\mu k_\mu) 
\big)
= \sum_\lambda 2 \lambda \amp{u}_\lambda(k,u) \overline{\amp{u}_\lambda(k,u)}\\
= S(L_u R_{\vec{n}}) 
\big(
\kappa + \gamma^5 (\gamma^\mu \tilde{k}_\mu)
\big)
S(L_u R_{\vec{n}})^{-1},
\end{multline}
\begin{multline}
\big(
\kappa - \gamma^5 (\gamma^\mu k_\mu) 
\big)
= \sum_\lambda 2 \lambda \amp{v}_\lambda(k,u) \overline{\amp{v}_\lambda(k,u)}\\
= S(L_u R_{\vec{n}}) 
\big(
\kappa - \gamma^5 (\gamma^\mu \tilde{k}_\mu)
\big)
S(L_u R_{\vec{n}})^{-1},
\end{multline}
Defining projectors
\begin{equation}
\Pi_+ = \tfrac{1}{2\kappa}
\big(
\kappa + \gamma^5 (\gamma^\mu k_\mu) 
\big),\quad
\Pi_- = \tfrac{1}{2\kappa}
\big(
\kappa - \gamma^5 (\gamma^\mu k_\mu) 
\big)
\end{equation}
we obtain
\begin{align}
&\Pi_+ + \Pi_- = I,&&\\
&\Pi_+ \amp{u}_\lambda(k,u) = \amp{u}_\lambda(k,u),
&& \Pi_+ \amp{v}_\lambda(k,u) = 0,\\
&\Pi_- \amp{v}_\lambda(k,u) = \amp{v}_\lambda(k,u),
&& \Pi_- \amp{u}_\lambda(k,u) = 0.
\end{align}
Defining operators
\begin{align}
\Pi_{\lambda+} = 2 \lambda \amp{u}_\lambda(k,u)
\overline{\amp{u}_\lambda(k,u)},\quad
\Pi_{\lambda-} = 2 \lambda \amp{v}_\lambda(k,u)
\overline{\amp{v}_\lambda(k,u)}
\end{align}
we get
\begin{align}
& \Pi_\pm = \sum_\lambda \Pi_{\lambda\pm},
&& \sum_\lambda(\Pi_{\lambda+}+\Pi_{\lambda-})=I,\\
&\Pi_{\lambda\pm} \Pi_{\sigma\pm} = 
2 \lambda \delta_{\lambda\sigma} \Pi_{\lambda\pm},
&& \Pi_{\lambda+} \Pi_{\sigma-} = 0.
\end{align}
and
\begin{align}
&\Pi_{\lambda+} \amp{u}_\sigma(k,u) =
\delta_{\lambda\sigma} \amp{u}_\sigma(k,u),
&& \Pi_{\lambda+} \amp{v}_\sigma(k,u) = 0,\\
&\Pi_{\lambda-} \amp{v}_\sigma(k,u) =
\delta_{\lambda\sigma} \amp{v}_\sigma(k,u),
&& \Pi_{\lambda-} \amp{u}_\sigma(k,u) = 0.
\end{align}

%\bibliography{quantum}


\end{document}